\documentclass[usenatbib,useAMS,usedcolumn]{mnras}
\usepackage{epsfig}
\usepackage{times,pdflscape}
\pdfoutput=1

\newcommand{\NHH}{\ensuremath{N_{\mathrm{H_{2}}}}}

\newcommand{\thco}{$^{13}$CO}

\newcommand{\etco}{C$^{18}$O}
\newcommand{\vel}{km\,s$^{-1}$}
\newcommand{\filname}{G350.5}
\newcommand{\filAname}{G350.5-N}
\newcommand{\filBname}{G350.5-S}
\newcommand{\imcoor}{$\alpha_{2000}=17^{\mathrm{h}}18^{\mathrm{m}}13\fs84,\ \delta_{2000}=-36\degr28\arcmin21\farcs5$}

\newcommand{\mline}{$M_{\rm line}$}
\newcommand{\msun}{$M_{\odot}$}

\newcommand{\cmcm}{cm$^{-2}$}
\newcommand{\egcite}{\citep[e.g.,][]}

\graphicspath{{./0figure/}}

\begin{document}

\title[ 
Large-scale periodic velocity oscillation in G350.54$+$0.69
] 
{ 
Large-scale periodic velocity oscillation in the filamentary cloud G350.54$+$0.69
}

\author[H.L. Liu et al.] {Hong-Li Liu\footnotemark$^{1,2,3}$, Amelia Stutz\footnotemark$^{3,4}$, Jing-Hua Yuan$^{5}$ \\
  $^1$ Chinese Academy of Sciences South America Center for Astronomy, China-Chile Joint Center for Astronomy, Camino El Observatorio \#1515, Las Condes, Santiago, Chile \\
  $^2$ Department of Physics, The Chinese University of Hong Kong, Shatin, NT, Hong Kong SAR \\
  $^3$ Departamento de Astronom\'ia, Universidad de Concepci\'on, Av. Esteban Iturra s/n, Distrito Universitario, 160-C, Chile \\ 
  $^4$ Max-Planck-Institute for Astronomy, K\"onigstuhl 17, 69117 Heidelberg, Germany\\ 
  $^5$ National Astronomical Observatories, Chinese Academy of Sciences, 20A Datun Road, Chaoyang District, Beijing 100012, China \\ 
}



\maketitle

\label{firstpage}

\begin{abstract} 
We use APEX mapping observations of \thco, and \etco~(2-1) to investigate the internal gas kinematics of the filamentary cloud G350.54$+$0.69, composed of the two distinct filaments \filAname\ and \filBname. G350.54$+$0.69 as a whole is supersonic and gravitationally bound.
We find a large-scale periodic velocity oscillation along the entire \filAname\ filament with a wavelength of $\sim 1.3$\,pc and an amplitude of $\sim 0.12$\,\vel.  Comparing with gravitational-instability induced core formation models, we conjecture that this periodic velocity oscillation could be driven by a combination of longitudinal gravitational instability and 
a large-scale periodic physical oscillation along the filament. The latter may be an example of an MHD transverse wave. This hypothesis can be tested with Zeeman and dust polarization measurements.
\end{abstract}

\begin{keywords}
  ISM: individual objects: G350.5 -- ISM: clouds -- ISM: structure -- ISM: molecules -- stars: formation -- infrared: ISM
\end{keywords}

\footnotetext[1]{E-mail: hongliliu2012@gmail.com}
\footnotetext[2]{E-mail: astutz@astro-udec.cl, stutz@mpia.de}

\section{Introduction}
\label{sec:intro}

Filamentary structures in the interstellar medium (ISM) have long been recognized \citep[e.g.,][]{sch79} but  
their role in the process of star formation has received focused attention
only recently thanks to long-wavelength Herschel data \egcite{mol10,and10}. With these data, recent
studies have demonstrated the ubiquity of the filaments in the cold ISM of our Milky Way \egcite{mol10,and10,stu15,stu18a}.
Meanwhile, analysis of Herschel data has revealed a close connection
between filamentary clouds and star formation \egcite{and10,and14,kon15,stu15,stu16,liu18}.  For instance,
most cores are detected in filamentary clouds \citep{and14,kon15,stu15}. 
Moreover, with gas velocity information, protostellar cores are found kinematically coupled to the dense filamentary
environments in Orion \citep{stu16}; that is, the protostellar cores have similar radial velocities as the gas filament.

The overarching goal of obtaining empirical observational information of the physical state of observed filaments 
has radically improved with the gradual availability and subsequent exploitation of gas radial velocities, which 
reveal internal kinematics of the filaments in great detail \egcite{hac11,tac14,hen14,beu15,taf15, hac17,hac18, lux18}.  
Indeed, the kinematics is the only way to probe not only where the mass is presently, but also most importantly how it is moving under the influence of various possible forces.  In short, the gas kinematics provide a powerful window into the physical processes at play in the conversion of gas mass into stellar mass in filaments.
Velocity gradients have been for example interpreted as being driven by gravity \egcite{hac11,tac14, hen16c, wil18, lux18, yua18}. For instance,
velocity gradients along the filament 
may be related to longitudinal collapse leading to core formation \egcite{hac11,kir13,per13,per14, tac14, hac17, wil18,lux18}, or those perpendicular to the filament 
may result from the filament rotation or radial collapse \egcite{rag12,kir13,beu15,wug18,liu18b}.

With the goal of understanding the link between dense cores and star formation  in filamentary clouds, we have investigated 
the filamentary cloud G350.54$+$0.69 (hereafter \filname) using {\it Herschel} continuum data (see Fig.\,1 of \citealt{liu18}, hereafter Paper\,I). \filname\ is a straight and isolated
in morphology and composed of two distinct filaments, \filAname\ in the north and \filBname\ in the south. G350.5-N(S) is $\sim 5.9$\,pc ($\sim 2.3$\,pc) long with a mass of 
$\sim 810$\,\msun~($\sim 110$\,\msun). The nine gravitationally bound dense cores associated with low-mass protostars suggest
 a site of ongoing low-mass star formation. In this work, we shift the focus to the gas kinematic properties. The nature of its simple morphology would be helpful in reducing the velocity ambiguities resulting from projection effects of potentially complex morphologies.

This paper is organized as follows: observations are described in Section\,2,  analysis results are presented
in Section\,3, the discussion is given in Section\,4 to a large-scale periodic velocity fluctuation along the main structure of the filament \filname, 
and a summary is put in Section\,5.

\section{Observations}
Observations of \thco, and \etco~(2-1) toward \filname\ were made simultaneously using the Atacama Pathfinder Experiment
(APEX\footnote{This publication is based on data acquired with the Atacama Pathfinder Experiment (APEX). APEX is a collaboration between the Max-Planck-Institut fur Radioastronomie, the European Southern Observatory, and the Onsala Space Observatory. }) 12--m telescope \citep{gus06} at Llano de Chajnantor (Chilean Andes) in the on-the-fly mode on September 24, 2017.
The observations were centered at \imcoor\ with a mapping size of $15\arcmin\times16\arcmin$, rotated by $50\degr$ relative to 
the RA decreasing direction. 
An  effective spectral resolution of 114\,KHz or 0.15\,\vel\ is reached at a tuned central frequency of 220\,GHz between
both for \thco~(2-1) and for \etco~(2-1). 
The angular resolution at this frequency is $\sim 28\arcsec$,  which corresponds to $\sim 0.2$\,pc at the distance of \filname, $1.38\pm0.34$\,pc (see Paper\,I). The reduced spectra finally present a typical rms value of 0.42\,K. More details on the 
data reduction can be found in Paper\,I.

\section{Results}
\label{sec:result}
\subsection{Spatial distribution of the gas filament}
\label{sec:gas:distr}
\begin{figure}
\centering
\includegraphics[width=3.4 in]{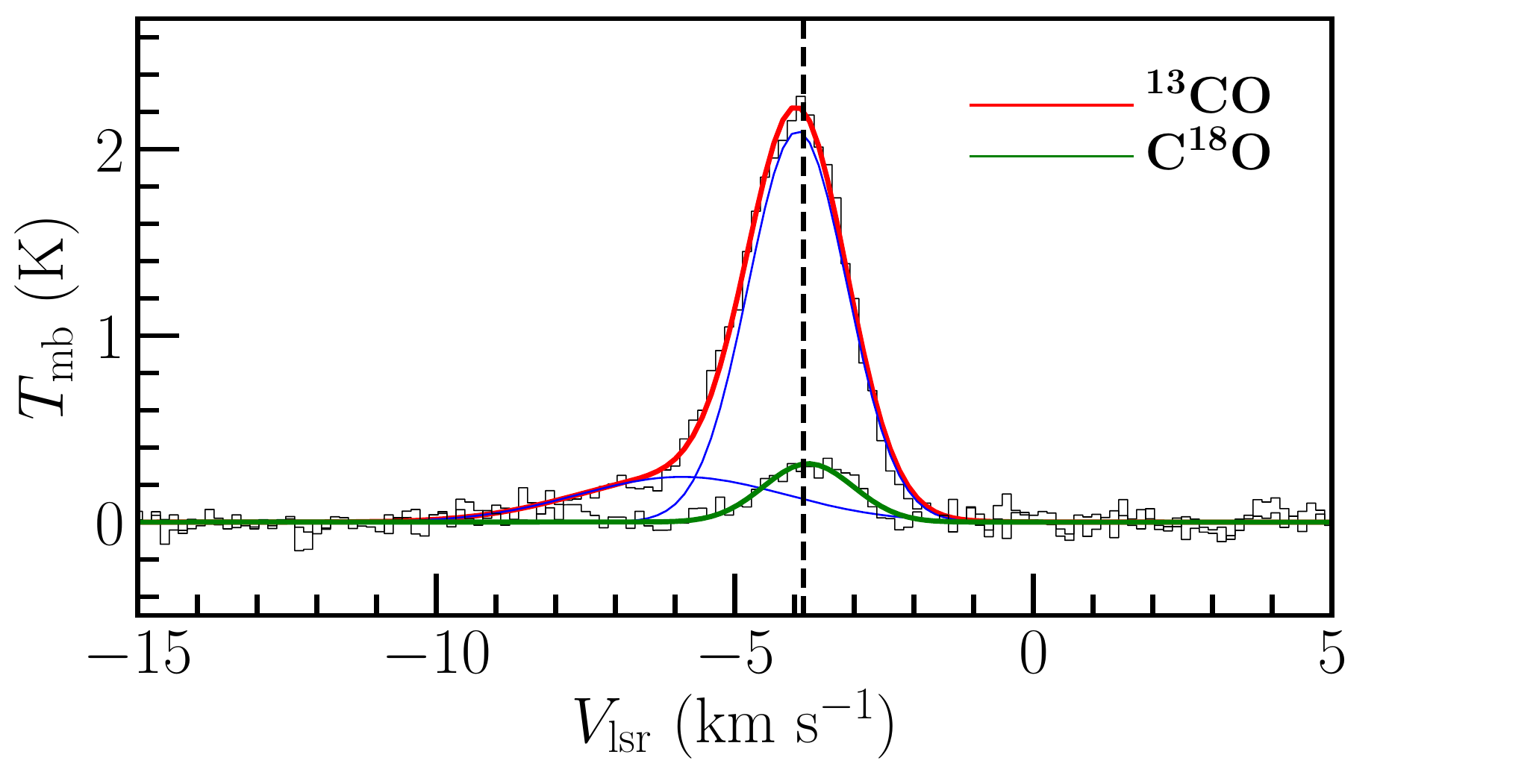}
\caption{Average spectra of \thco, and \etco~(2-1) over the \filname\ filamentary cloud. The black color stands for the observed spectrum.
The red line is the Gaussian fitting to the spectrum of \thco, which is a sum of the two Gaussian components (blue lines), the weak-emission one from $-9.0$ to
$5.7$\,\vel, and the  major one from $-5.7$ to $-2$\,\vel. The green line is the single-component Gaussian fitting to the spectrum of \etco, which matches the major 
component of the \thco\ spectrum. The vertical dashed black line represents a systematic velocity of $-3.9$\,\vel.
} 
\label{fig-spec_aver}
\end{figure}

Before addressing the kinematic structure of \filname, we first describe the spatial distribution of molecular emission. 
Figure\,\ref{fig-spec_aver} presents the average spectra of \thco, and \etco~(2-1) over the whole \filname\ system. 
The \thco\ profile has two velocity components, one from $-9.0$ to
$5.7$\,\vel\ (hereafter weak-emission component) and the other from $-5.7$ to $-2$\,\vel\ (hereafter major component); 
\etco\ has a single-peak profile corresponding to the major component of \thco.
The systemic velocity is determined to be  $-3.9$\,\vel\ from the average spectrum of \etco~(2-1).

\begin{figure*}
\centering
  \includegraphics[width=6.8 in]{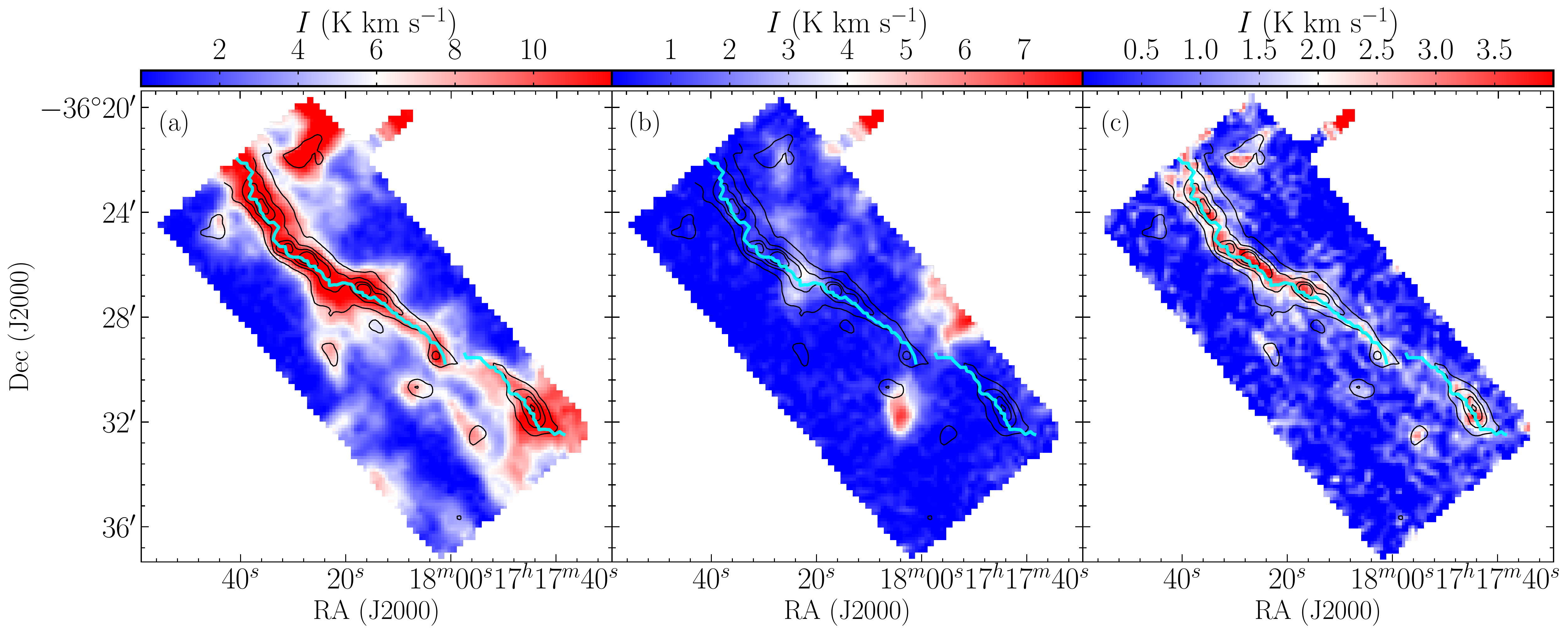}
\caption{(a--b:) Velocity-integrated intensity maps of the two velocity components of \thco~(2-1) toward \filname, the major one from $-5.7$~to~$-2$\,\vel, and the weak-emission one
from $-9$~to~$-5.7$\,\vel.  (c:) Velocity-integrated intensity of \etco~(2-1), integrated over $-7$ to $-1$\,\vel.  The pixel size of the three maps is reduced to 7$\arcsec$, a quarter of the beam size, for a better visualization.  The cyan curves in all panels represent the spines of the two discontinuous filaments as identified in Paper\,I and the black contours stand for the $N_{\rm H_2}$ column density (see Paper\,I), starting from $1.8\times10^{22}$\,\cmcm\ with a step of $0.6\times10^{22}$\,\cmcm\ ($\sigma=0.04\times10^{22}$\,\cmcm).
}
\label{fig-13CO_emission}
\end{figure*}

Figure\,\ref{fig-13CO_emission} shows the velocity-integrated intensity maps of  \thco, and \etco~(2-1). The spatial distribution of the major component of \thco\ (see Fig.\,\ref{fig-13CO_emission}a) matches the column density (\NHH) distribution (black contours, derived from Herschel data in Paper\,I through the pixel-wise spectral energy distribution fitting, e.g., \citealt{liu16,liu17}), showing two discontinuous filaments as indicated by the two cyan curves.
 In addition, some small-scale structures (i.e., filament `branches') stretch perpendicular to the main structure of the cloud \filname, which
coincide with the perpendicular dust striations surrounding the filament as observed in Herschel continuum
images (see Paper\,I).  Emission of the major component of \thco~(2-1) has a \etco\ counterpart  (in Fig.\,\ref{fig-13CO_emission}c), but the latter is more narrowly distributed than the former. 
In contrast, the weak-emission component of \thco~(2-1) in Fig.\,\ref{fig-13CO_emission}b has no detectable \etco\ counterpart. However, it can be seen that the two components of  \thco\ are indeed associated with each other. In the south and center of the filament, the two most prominent clumps in the intensity map of the \thco\ weak-emission component can also be found in that of the major component. This is supportive of the association between the weak-emission and major components.

\subsection{Velocity field along the filament}
\label{sec:vel:cent}
\begin{figure}
\centering
\includegraphics[width=3.4 in]{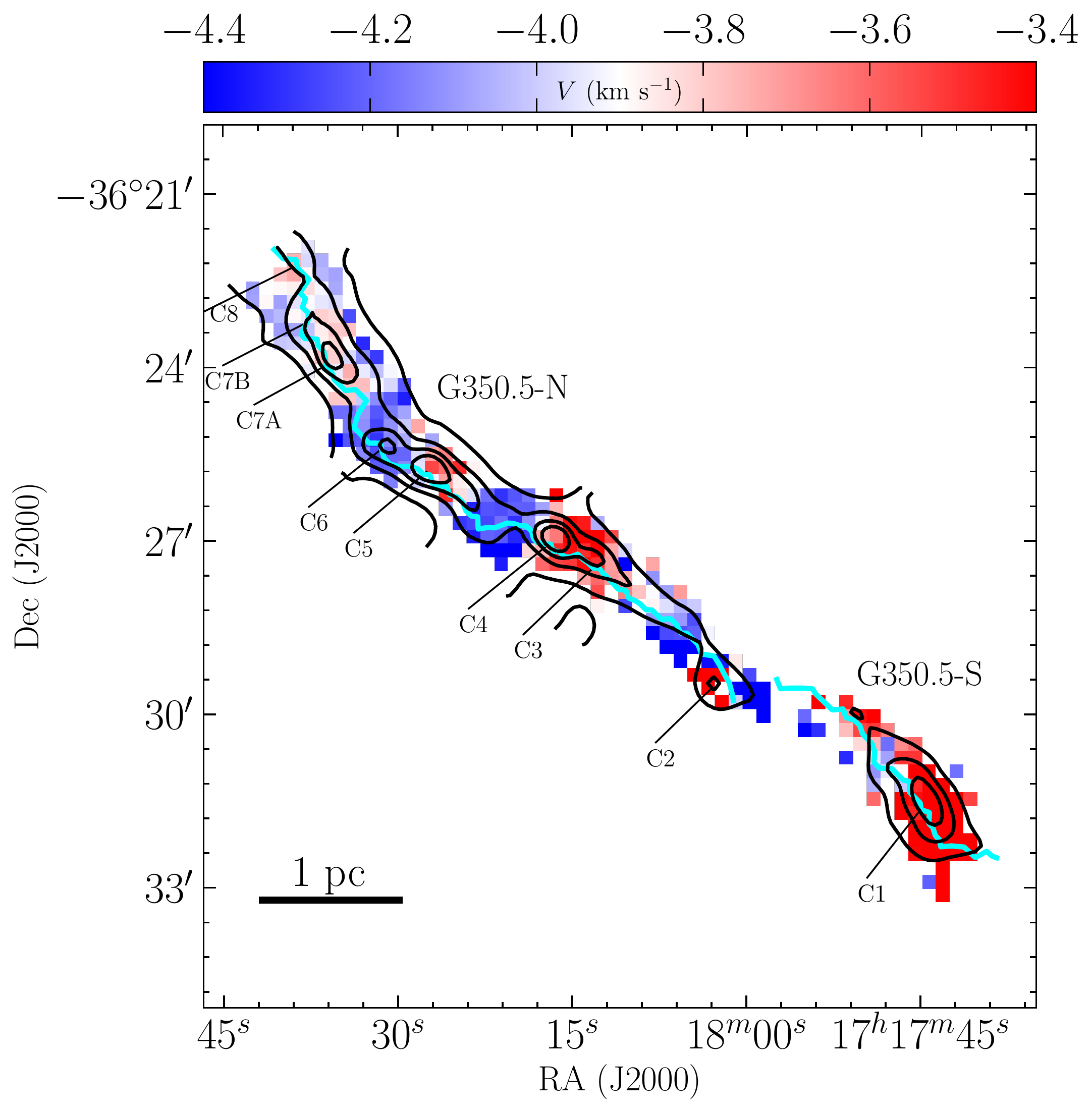}
\caption{Velocity centroid map of \etco~(2-1) for the filamentary cloud \filname. Only the main filament is shown
because of sufficient detection of \etco~(2-1) therein. The pixel size of the map is 14$\arcsec$, half of the beam size, as recommended by the APEX observatory.
All pixels correspond to detection greater than  $3\sigma$ (i.e., $\sigma=0.4$\,K\,\vel). 
The red and blue color scales can be regarded as the red and blue-shifted velocities with respect to 
the systematic velocity $-3.9$\,\vel.  A large-scale periodic velocity fluctuation appears almost 
along the main structure. The black contours represent the $N_{\rm H_2}$ column density, starting from
$1.8\times10^{22}$\,\cmcm\ with a step of $0.6\times10^{22}$\,\cmcm. 
The dense cores identified in Paper\,I are also labelled as C1--C8.
}
\label{fig-C18O_emission_vel}
\end{figure}

To investigate the large-scale kinematics, we show  the velocity centroid map of \etco~(2-1) of the cloud \filname\ in Fig.\,\ref{fig-C18O_emission_vel}.
Velocity information is shown only in the main structure of the filament, beyond which \etco~(2-1) emission is too noisy ($<2\sigma$), overlaid with 
the \NHH\ contours (Paper\,I) for comparison. Strikingly, a large-scale periodic velocity fluctuation appears along the main structure, which is $\sim8$\,pc long (Paper\,I).
The nature of the large-scale and periodic signal suggests that the observed velocity fluctuation is real, otherwise, this periodicity will be hardly maintained on a large scale if it happens by chance.
Specifically,  the red and blue colors in Fig.\,\ref{fig-C18O_emission_vel} represent the red and 
blue-shifted velocities relative to the systemic velocity of $-3.9$\,\vel. Comparing the \NHH\ contours with the velocity centroid map,
we find a spatial correspondence between the velocity extremes and the density enhancements for some dense cores (i.e., C2, C5, C6, C7a,b, and C8).
The possible origins of this correspondence will be discussed
further in Sect.\,\ref{sec:disc}. We do not present here the velocity centroid map of \thco~(2-1), which does not show the periodic fluctuation as in \etco~(2-1) emission.
Since \thco\ tends to trace more extended emission than \etco, the velocity characteristics revealed by both species are not necessarily the same.

\subsection{Velocity dispersion along the filament measured from both \thco\ and \etco}
\begin{figure*}
\centering
\includegraphics[width=3.4 in]{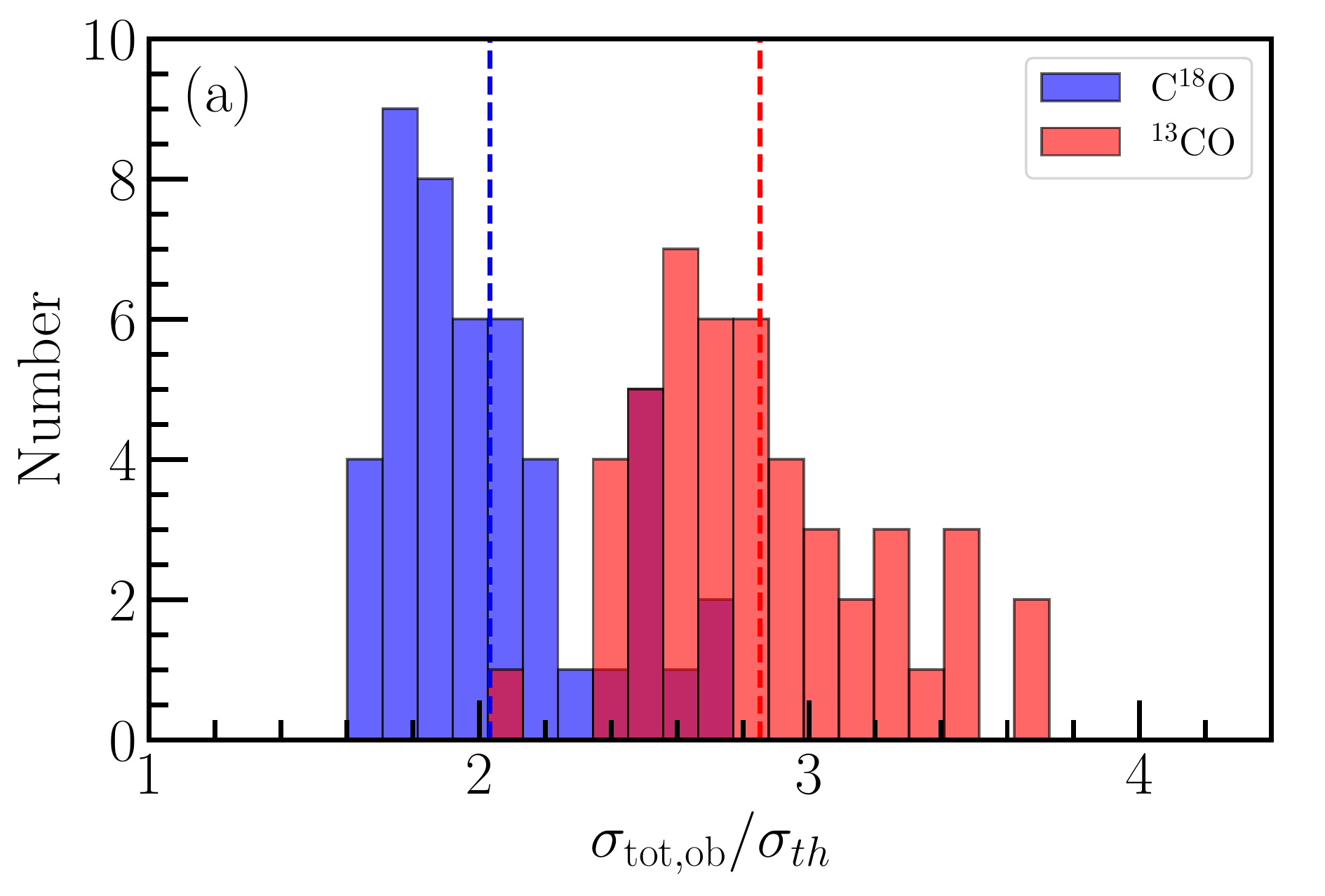}
\includegraphics[width=3.4 in]{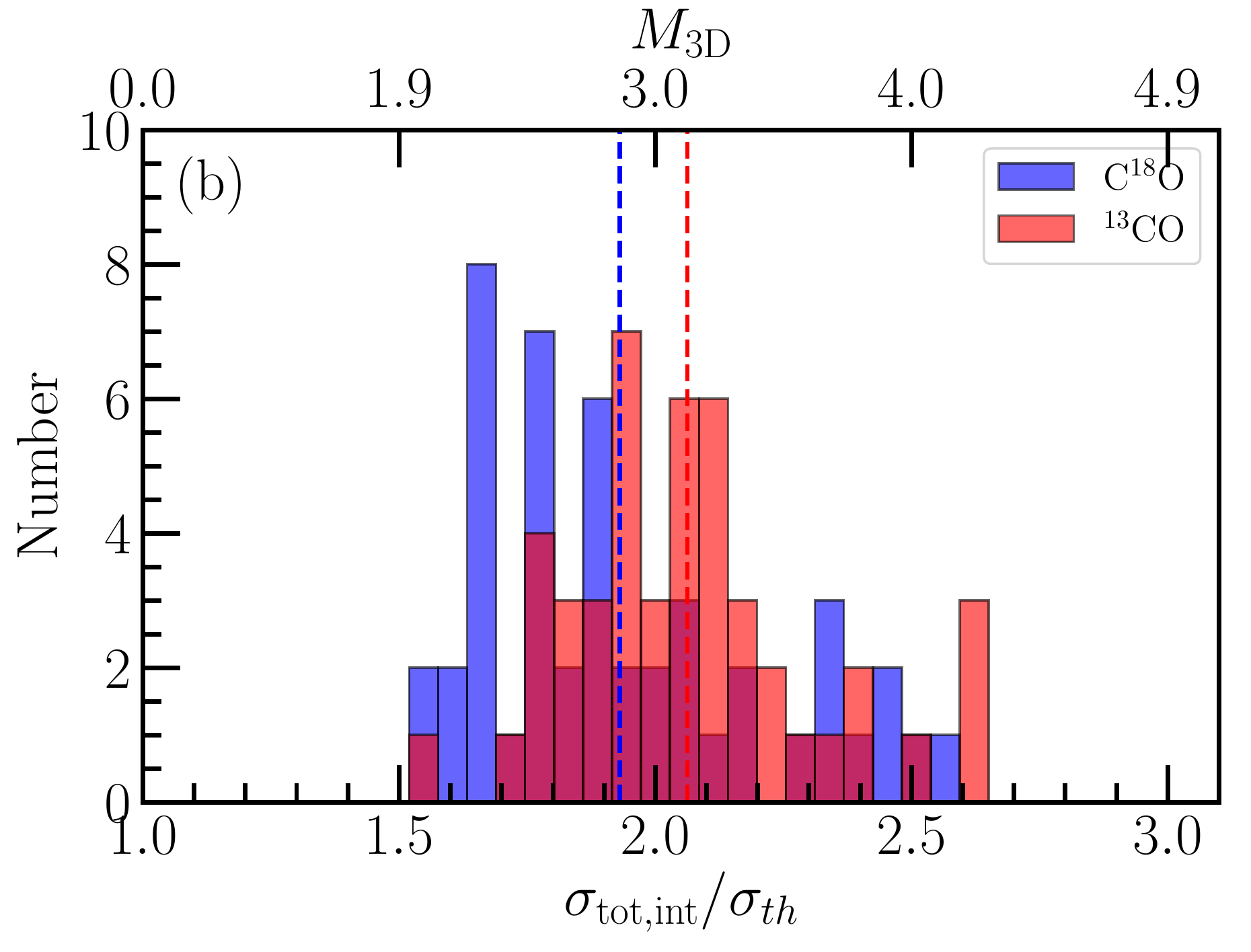}
\caption{Histogram of the total velocity dispersion measured from both \thco\ (red) and \etco~(2-1) (blue) for \filname. In panel\,(a) is the observed velocity dispersion.
The mean values in the observed velocity dispersions measured from  \thco\ and \etco\ are $2.9\pm0.4\,\sigma_{\rm th}$ (red dashed line), 
and $2.0\pm0.3\,\sigma_{\rm th}$ (blue dashed line), respectively. 
In panel\,(b) is the opacity-corrected (intrinsic) velocity dispersion. The mean values are $2.1\pm0.2\,\sigma_{\rm th}$ from \thco\ (red dashed line), 
and  $1.9\pm0.3\,\sigma_{\rm th}$ from \etco\ (blue dashed line). 
The different distributions of the velocity dispersions of \thco\ before and after the opacity correction demonstrate that the line opacity can cause the broadening of 
the intrinsic velocity dispersion. The intrinsic velocity dispersion distributions indicate that the  overall turbulent motions of the filament G350.5 are supersonic (i.e., the 3D Mach 
number follows $M_{3D} >2$).
}
\label{fig-vel_disp}
\end{figure*}

The observed velocity dispersion ($\sigma_{\rm obs}$) is measured from both \thco\ and \etco~(2-1) in
the main structure of the filament where both species are well-detected. In principle, $\sigma_{\rm obs}$ 
can be estimated from the second-order moment of the PPV data cube. However, this method will cause additional uncertainties due to the simple treatment of 
multiple-velocity components as one. To better estimate $\sigma_{\rm obs}$ especially from \thco~2-1, we define 47 positions along the ridgeline of the filament, 41 in \filAname\ and 6 in \filBname. 
Neighbour positions are separated by one pixel, half of the beam size of both \thco\ and \etco\ maps. 
$\sigma_{\rm obs}$ is then obtained from the Gaussian fitting to the spectra of both \thco, and \etco~(2-1) at the selected 47 positions.
The fitting plots are shown in Fig.\,\ref{fig-guassA}. 

The velocity dispersion combines the contributions from both thermal and non-thermal gas motions in turbulent clouds. However, their contributions can be overestimated due to the line broadening
by optical depth.  Such overestimate can be expressed analytically as a function of the optical depth \citep{phi79}:
\begin{equation}\label{eq:broad}
\sigma_{\rm obs} =\frac{1}{\sqrt{ln 2}}\left[ln \left( \frac{\tau}{ln\left(\frac{2}{e^{-\tau}+1} \right) } \right) \right]^{1/2} \sigma_{\rm int},
\end{equation}
where $\sigma_{\rm int}$ is the intrinsic velocity dispersion unaffected by optical depth $\tau$ (see Appendix\,\ref{sec:opt:dep}). 
For comparison, we calculate the gas dispersions (i.e., non-thermal dispersion $\sigma_{\rm NT}$, and the total dispersion $\sigma_{\rm tot}$)  based on the observed, and intrinsic velocity dispersions measured from both \thco\ and \etco, respectively, as follows:
\begin{equation}\label{eq:vel:disp}
\begin{array}{rrr}
\sigma_{\rm NT} = \sqrt{\sigma_{\rm obs/int}^2-\frac{kT_{\rm k}}{m}}, \\   
\sigma_{\rm th} = \sqrt{\frac{kT_{\rm k}}{ \mu_p m_{\rm H} }}, \\
\sigma_{\rm tot} = \sqrt{\sigma_{\rm NT}^2+\sigma_{\rm th}^2}, \\
\end{array}
\end{equation}
where $\sigma_{\rm obs/int}$ can be replaced either with $\sigma_{\rm obs}$ or with $\sigma_{\rm int}$ (see Eq.\,\ref{eq:broad}), $k$ is  Boltzmann’s constant, $T_{\rm k}$ is the gas kinetic temperature, $m$ is the mass of the
two species,  $m_{\rm H}$ is the mass of atomic hydrogen, and $\mu_p=2.33$ is the mean molecular weight per free particle for an abundance ratio
$N(H)/N(He)=10$ and a negligible admixture of metals (e.g., \citealt{kau08}). 
The gas kinetic temperature here was assumed to be equal to the average dust temperature ($T_{\rm d} \sim18 \,K$) of the filament instead of the excitation temperature of
each species since the latter is found to be $\sim6$\,K on average lower than the former, indicating that the two species might 
not be fully thermalized in the filament (see Appendix\,\ref{sec:ex:temp} for the calculation of the excitation temperature).
The assumed $T_{\rm k}$ gives rise to the thermal sound speed  of molecular gas $\sigma_{\rm th}\sim0.25$\,\vel.

Figure\,\ref{fig-vel_disp} presents the histograms of both the observed ($\sigma_{\rm tot,obs}$ in panel a) and intrinsic ($\sigma_{\rm tot,int}$ in panel b) total dispersions of gas
for the entire filament represented by 41 positions in \filAname\ and 6 in \filBname. The intrinsic gas dispersion is overall smaller than the observed one without the optical-depth correction, indicating that the optical depth correction is important, especially for \thco.   In view of this, we analyse only the intrinsic gas dispersion (in Fig.\,\ref{fig-vel_disp}b) in what follows. 
The average total dispersions are $2.1\pm0.2\,\sigma_{\rm th}$, and $1.9\pm0.3\,\sigma_{\rm th}$, measured from \thco, and \etco, respectively.
These values indicate that the
filament \filname\ as a whole is supersonic with a mach number of $M_{3D} > 2$ (see Fig.\,\ref{fig-vel_disp}b), where 
the 3D Mach number is $M_{3D}= \sqrt{3} \sigma_{\rm NT,int}/\sigma_{\rm th}$ given the 1D measurement $\sigma_{\rm NT,int}$, and assuming isotropic turbulence
 in three dimensions \egcite{kai17b}.

\subsection{Virial analysis along the filament}
\label{sec:vel_fluct}
\begin{figure*}
\centering
\includegraphics[width=3.2 in]{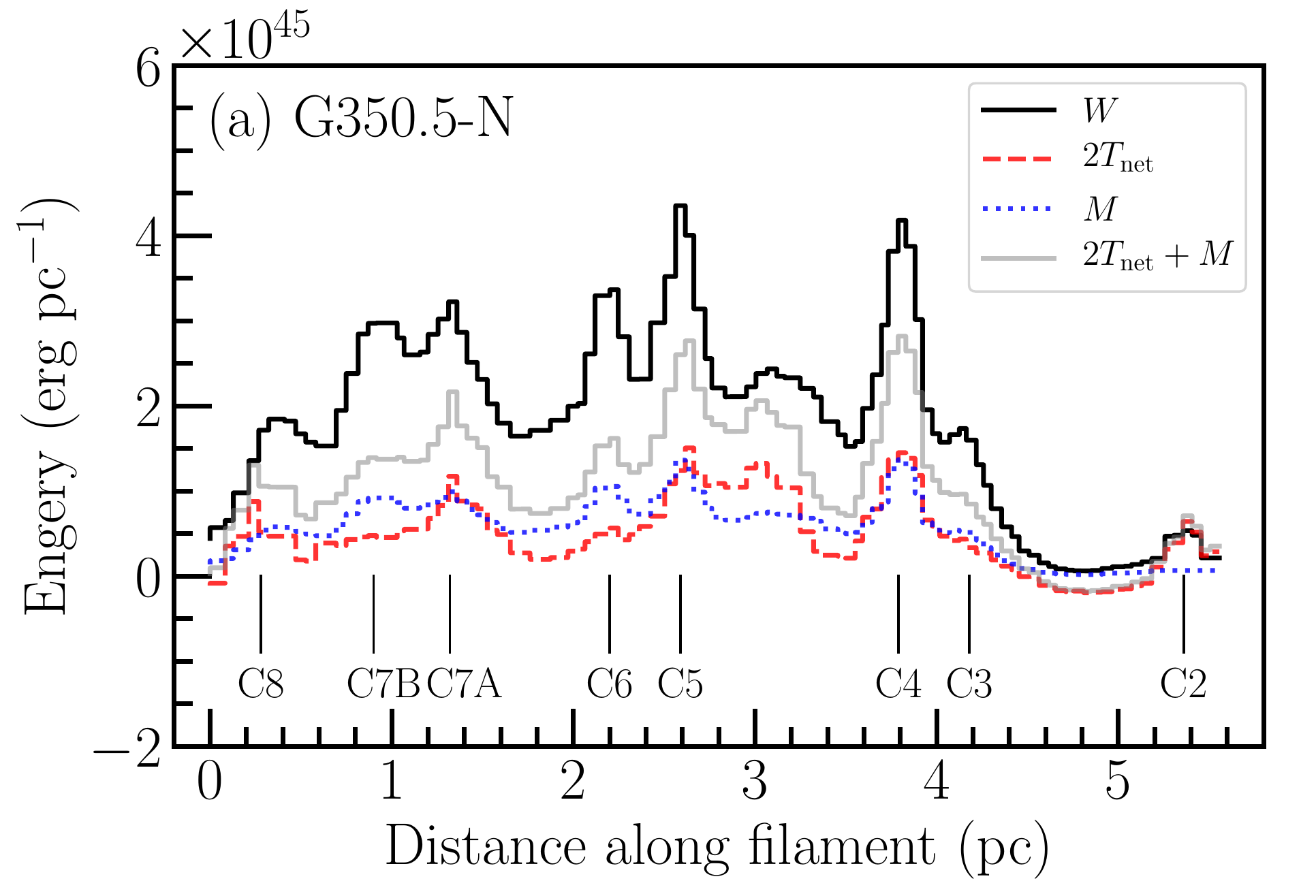}
\includegraphics[width=3.2 in]{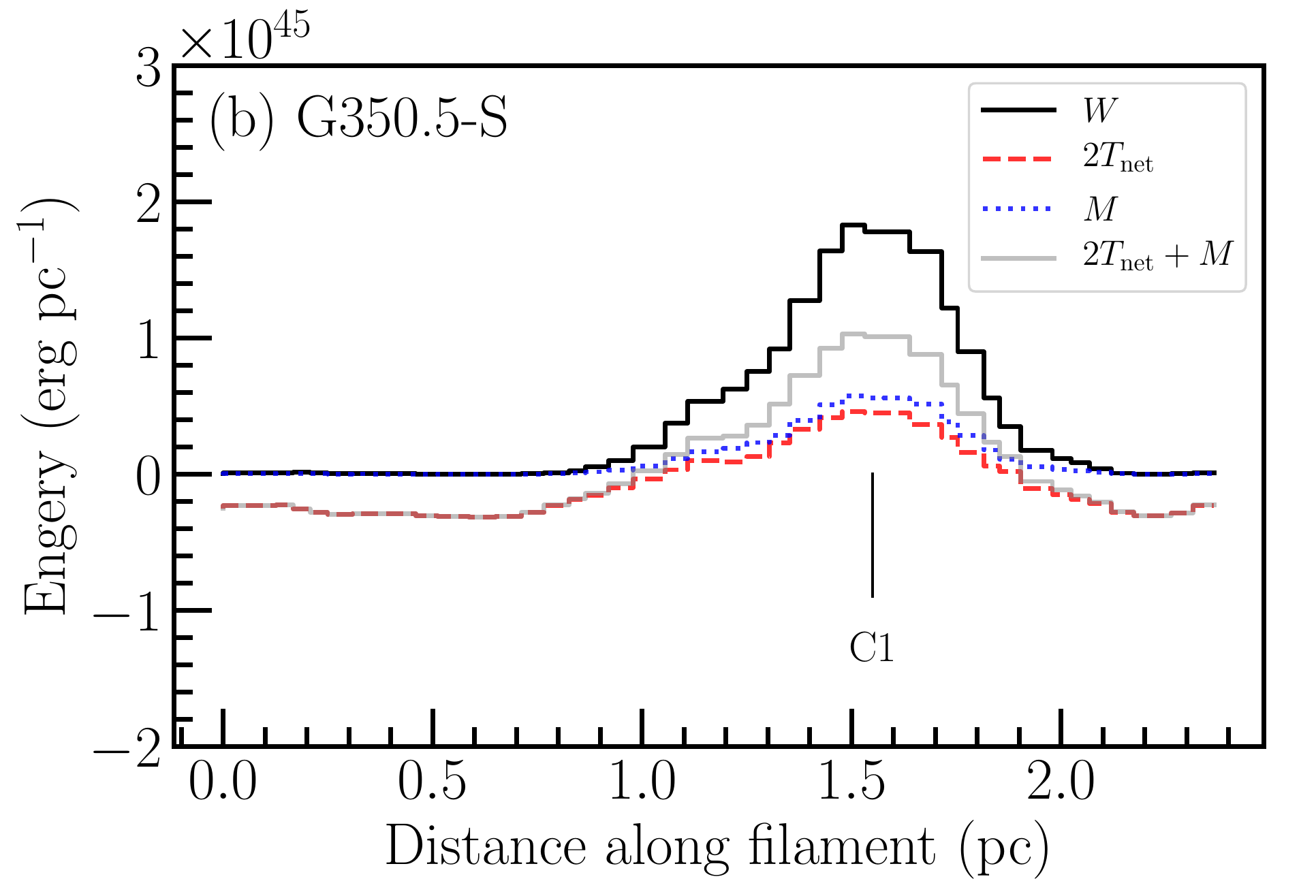}
\caption{Virial analysis through the comparisons between turbulent ($T$), magnetic energies ($M$) and gravitational potential ($W$) for \filAname\ (in panel\,a), and 
for \filBname\ (in panel\,b).  Several peaks are related to dense cores (also see Fig.\,\ref{fig-C18O_emission_vel}) as indicated by vertical lines. The comparison shows that
the gravity dominates over both turbulence and magnetic fields, and is a major driver to the global fragmentation of G350.5 from clouds to dense cores.
}
\label{fig-virial}
\end{figure*}
The complex interactions between turbulence, gravity and magnetic fields happen everywhere in  molecular clouds and regulate star formation therein.
We can describe these interactions  through the virial theorem.
According to \citet{fie00}, the virial equation of self-gravitating, magnetized turbulent filamentary clouds can be written in the form:
\begin{equation}\label{eq:luminosity}\label{eq:plummer}
2T_{\rm net}+W+M = 0.
\end{equation}
Therefore, the virial equilibrium  of the clouds depends on the competition between the net kinetic energy ($T_{\rm net}$), magnetic energy ($M$), and gravitational potential ($W$).
$T_{\rm net}$ accounts for the difference between the internal ($T_{\rm int}$) and external ($T_{\rm ext}$) turbulent energy, assuming that molecular clouds are confined by the external pressure rather than completely isolated. Note that all energies here are measured per unit length.

The internal kinetic energy $T_{\rm int}$ is calculated as:
\begin{equation}
T_{\rm int} = \frac{1}{2}\ M_{\rm line}\ \sigma^2_{\rm tot,int},
\end{equation}
where \mline\ is the line mass along the filament, which can be obtained in Fig.\,3 of Paper\,I. 
While $\sigma_{\rm tot,int}$ measured from \thco\, and \etco~(2-1) are similar,
the one from \thco~(2-1) was finally adopted in the calculation since \thco\ emission tends to trace larger-scale gas. 
 The external turbulent energy $T_{\rm ext}$ can be expressed as a function of the external pressure $P_{\rm ext}$:
\begin{equation}
T_{\rm ext} =  {\it k} P_{\rm ext}\ \pi R_{\rm fil}^2,
\end{equation}
where {\it k} is the Boltzmann constant and $R_{\rm fil}$ is the filament radius, $\sim 0.5$\,pc as measured in Paper\,I. A conservative value of $P_{\rm ext}=5\times10^4$\,K\,cm$^{-3}$ is assumed, 
which is in the range between $10^4$\,K\,cm$^{-3}$ for the general ISM \citep{chr89} and $10^5$\,K\,cm$^{-3}$ for several molecular clouds associated 
with HI complexes \citep{bou90}.

The gravitational potential energy $W$ is derived from the line mass of the filament:
\begin{equation}
W = -M_{\rm line}^2 G,
\end{equation}
where $G$ is the gravitational constant.
The magnetic energy  $M$ is written as:
\begin{equation}
M = \frac{B^2}{2\mu_0} \pi R_{\rm fil}^2,
\end{equation}
where $B$ is the magnetic field strength and  ${\mu_0}$ is the permeability of free space.
$B$ is estimated following the empirical linear relationship between the field strength and
gas column density. This relationship was summarized by \citet{cru12} from Zeeman measurements in the form:
\begin{equation}\label{eq:luminosity}\label{eq:plummer}
B = C \times N_{H}\, 10^{-21}\,\mu G,
\end{equation}
where $C$ is a constant, and $N_{\rm H}$ is the column density of atomic hydrogen gas. This constant was given to be 3.8 by \citet{cru12} under the assumption of magnetic critical condition and a spherical shape of clouds, and improved to be 1.9 by \citet{li 14} assuming a sheet-like shape.
The mean column density of the cloud \filname, $N_{\rm H_2} = 8\times10^{21}$\,\cmcm\ (Paper\,I), indicates
an average B-field of $16$\,$\mu$G. Even though this estimate of the B-field strength is very indirect, it can give us some insight into the role of the B-field versus gravity (see below).

Figure\,\ref{fig-virial} presents the comparison between turbulent ($T$), gravitational ($W$) and magnetic ($M$) energies per unit length along the main structure of
the two filaments \filAname, and \filBname.  It can be seen that $W$ (black curve) is greater than the sum of $T_{\rm net}$ and $M$ (light grey curve) 
along the main structure of both \filAname\ and \filBname\ with respective mean ratios of $W/(T_{\rm net}+M)=2.0\pm1.6$, and $2.5\pm1.8$. This trend is in particular evident
for the dense cores which correspond to the peaks in the distributions (see Fig.\,\ref{fig-virial}). This suggests that gravity dominates over both turbulence and
 magnetic fields, and is a  major driver to the global fragmentation
of \filname\ from clouds to dense cores. This is consistent with  our previous analysis (Paper\,I) showing that \filname\ could have undergone radial collapse and 
fragmentation into distinct small-scale dense cores. Moreover, one can see that only turbulence without the aid of the B-field is not able to counteract gravity, suggesting the critical role of B-field in regulating molecular clouds against gravity. This is in agreement with existing observations showing that magnetic fields are important in filamentary clouds  \egcite{li 13,li 14,li 15, stu16, con16}. Future more accurate B-field strength measurements will provide direct constraints on the role of the magnetic fields in \filname.

\section{Discussion: Large-scale velocity oscillations along the filament}
\label{sec:disc}
\begin{figure*}
\centering
\includegraphics[width=6.0 in]{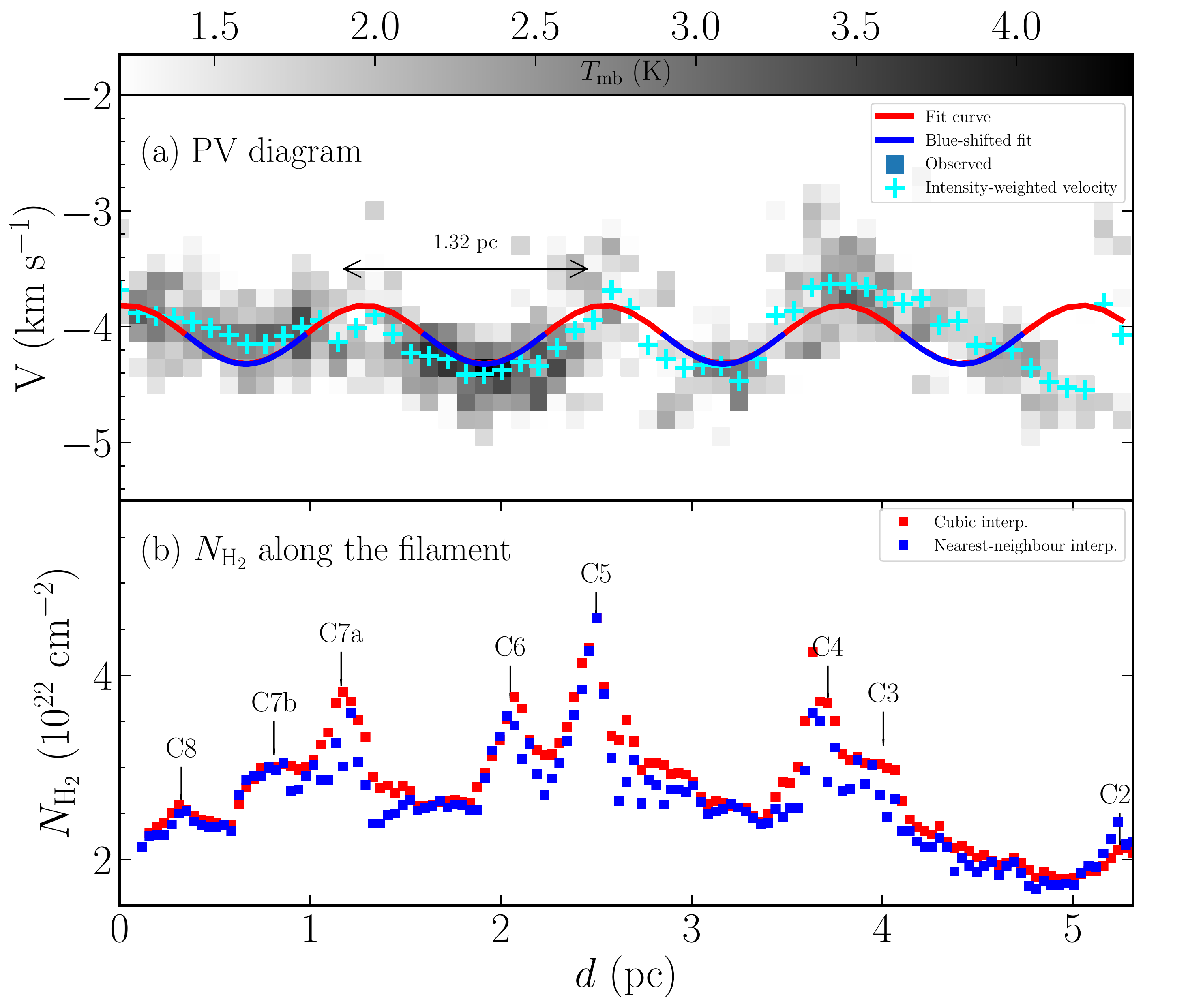}
\caption{(a-b:) Velocity and column density distributions extracted along the spine of the filament \filAname. 
(a): Position-velocity diagram of \etco~(2-1). The detections greater than $3\sigma$ ($\sigma=0.42$\,K) are shown in grey-scale within the velocity channels from $-5.0$ to $-2.5$\,\vel. 
The cyan crosses are the weighted average velocities by brightness temperatures, and fitted with  a sine function with a wavelength of $\sim 1.3$\,pc and an amplitude of $\sim 0.12$\,\vel. Literally, the blue 
color curves stand for the blue-shifted velocities with respect to the systematic velocity of $-3.9$\,\vel, and the red ones for the red-shifted velocities.
(b:) Density distribution. Density peaks related to dense cores are indicated, i.e., C2 to C8, which are extracted from Fig.\,\ref{fig-C18O_emission_vel}. The velocity and column 
density distributions do not show a good one-to-one correspondence between one another.
}
\label{fig-pvNH2}
\end{figure*}

\begin{figure*}
\centering
\includegraphics[width=7.0 in]{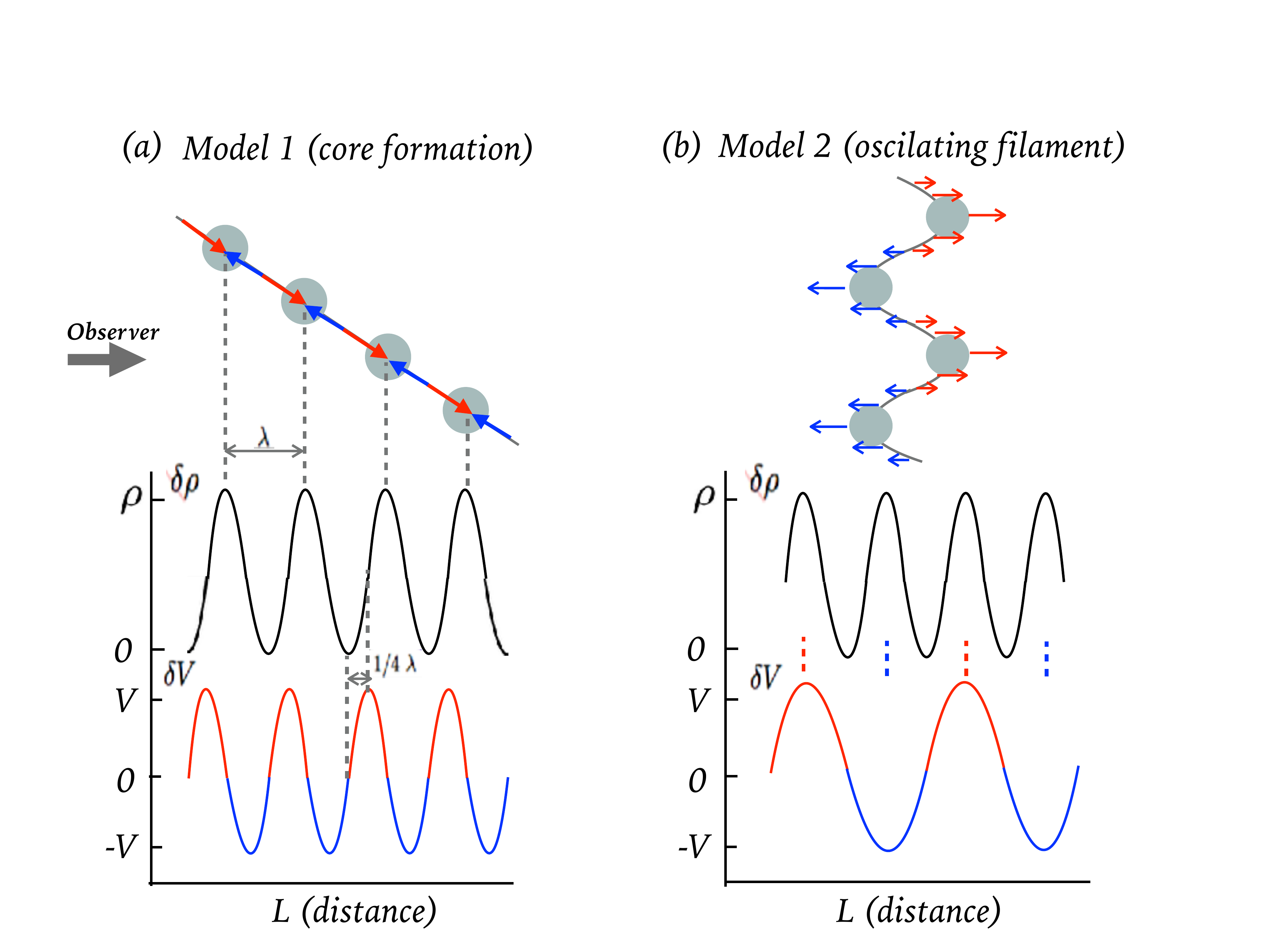}
\caption{(a:) schematic representation of GI-driven core formation along a filament, which is adapted from Fig.\,12 of \citet{hac11}.
Due to GI, the motions of core-forming gas converge towards the core centers where the density enhances 
but the relative motions become static with respect to the systematic velocity. In this scenario, assuming that both  density ($\delta \rho$) and velocity ($\delta V$) follow sinusoidal distributions, 
a $\lambda/4$ phase shift between them would be expected. (b): schematic representation of a physical oscillating filament (e.g., induced by an MHD wave). The filament moves toward and away from  
an observer. In this scenario, the correspondence between the velocity extremes (i.e., red/blue-shifted velocities) and the density enhancements is expected if the filament (density) oscillates sinusoidally. The red and blue colors indicate the red, and blue-shifted velocities (motions) with respect to an observer. Note that  the physical intensities of $\delta \rho$ and $\delta V$ are ignored in the schematic demonstration.
}
\label{fig-models}
\end{figure*}

As mentioned in Sect.\,\ref{sec:vel:cent}, a large-scale periodic velocity oscillation is found  along the entire filament. 
This oscillation feature may be related to kinematics of either core formation or a  large-scale oscillation
(e.g., wave-like perturbations triggered by material accretion flows onto the filament or  a standing wave \citet{stu18b}). In what follows, we will make an attempt to 
investigate the nature of the observed large-scale velocity oscillation in \filAname. Since \filBname\ is disconnected to \filAname\ in Fig.\,\ref{fig-C18O_emission_vel} and \filBname\ itself has insufficient detection in the diffuse region, the nature
of the velocity field of \filBname\ requires detailed investigation with sensitive and higher-resolution observations. To better visualize the
velocity oscillation along \filAname, we make the position-velocity (PV) diagram shown in Fig.\,\ref{fig-pvNH2}a. 
The color scale is the main-beam temperature with  at least $3\sigma$ ($\sigma=0.42$\,K) detection within $-5.0$ to $-2.5$\,\vel. 

To highlight the velocity fluctuation feature, we calculate the average velocities weighted by the main-beam temperature, as shown in cyan crosses.
In Fig.\,\ref{fig-pvNH2}a, the velocity oscillation behaves periodically along the filament. Particularly, it can be fit  with a sine function with a  wavelength of $\sim 1.3$\,pc and an amplitude of $\sim 0.12$\,\vel.

\subsection{Core formation through longitudinal gravitational instability?}
\label{sec:core:form:grav:contr}

In Paper\,I, nine identified cores are found to be distributed almost periodically along the 
entire filamentary cloud \filname\ and the average projected separations between them are measured to be  $\sim 0.9$\,pc.
This separation is consistent with the prediction ($\sim 1.1$\,pc in Paper\,I) 
by the ``sausage'' (\textbf{gravitational}) instability \citep{cha53,nag87}, and comparable to the wavelength of the periodic velocity signature 
($\sim 1.32$\,pc, see above) inferred from the periodic velocity distribution as well. Therefore, the periodic velocity fluctuation may be 
associated with the kinematics of filament fragmentation into cores through gravitational instability (GI hereafter). 

Actually, this periodic-type velocity fluctuation was already predicted in the analytic models dedicated to GI-induced core formation 
in a uniform, incompressible filament threaded by a purely poloidal magnetic field  \egcite{cha53}.
In simulations, the GI is generally represented with 
redistributions of the gas in the filament via motions that have a dominant velocity component parallel to the cylinder axis due to longitudinal 
gravitational contraction at least during the first stages of evolution \egcite{nak93,fie00b}. The gas redistribution processes can be summarized in
the schematic model of core formation as seen in Fig.\,\ref{fig-models}a, where the motions of core-forming gas converge towards the core center along the filament,
leading to the density enhancements peaking at a position of vanishing velocity. Assuming that both density and velocity
perturbations (oscillations) are sinusoidal (periodic), a $\lambda/4$ shift between them can be expected (\citealt{hac11} for 
more details).  The similar pattern of velocity oscillation and its association with the density distribution have been reported in Taurus/L1517 \citep{hac11}.

Comparing with the scenario of GI-induced core formation models, we do not observe a $\lambda/4$ ($\sim 0.33$\,pc) phase shift between
the velocity and density distributions (see Fig.\,\ref{fig-pvNH2}).
In addition, the predicted vanishing of velocity (see above) is not observed in most of the density enhancements cores (i.e., C2, C5, C6, C7a,b, and C8),
and instead the velocity extremes coincide spatially with the density enhancements.
Note that the above-mentioned models are rather idealised, and only designed  for a straight filament. For example,
if a filament were randomly kinked/curved on small scales, neither the systematic phase shift between the velocity and density distributions or 
the  vanishing of velocity at the position of cores would be expected in the core formation process (see scenario\,1 in Fig.\,12 of \citealt{hen14}).
However, we believe that the possibility of a randomly kinked structure is very low in \filAname\ since (random) small-scale kinks would not maintain a 
large-scale coherent, periodic oscillation. On the other hand, a regular but oscillating geometry driven by some physical mechanism could be possible 
(e.g., \citealt{gri17}, see model\,2 in Fig.\,\ref{fig-models}, and Sect.\,\ref{sec:mhd} for more discussions).

Moreover, upon inspection of  Fig.\,\ref{fig-C18O_emission_vel}, we can group all dense cores into six main mass-accumulation clumps, i.e., 
C1 [red], C2 [red], C3+4 [red], C5+6 [red+blue], C7a+b [red+blue], and C8 [red]. They are almost periodically separated, which is demonstrated to be 
a result of filament fragmentation through GI (see above, and Paper\,I). Three of these clumps might have undergone further fragmentation
due to GI or Jeans collapse \egcite{kai17}, leading to the observed multiplicity (i.e., from clumps to cores; C3+4 to C3 and C4, C5+6 to C5 and C6, and C7a+b to C7a and C7b). 
Such clump-scale fragmentation may influence 
the small-scale (inter-core) velocity distribution within the clumps. As a result, red and blue-shifted velocities would be expected on either side of dense cores in
a straight filament.  Assuming that \filAname\ is straight to some extent (see above), no red and blue-shifted velocities appearing on
either side of the dense cores within two of the three clumps (i.e., C5+6, and C7a+b), therefore, imply that  
the clump-scale fragmentation (if any)  would not significantly affect  the large-scale periodic velocity pattern.

In addition, the average amplitude of the velocity oscillations is measured to be $\sim 0.12$\,\vel. This value is
 around three times greater than that observed in Taurus/L1517 \citep{hac11}. This difference could depend on the mass of
 dense cores and the inclination of the filament. Actually, the average mass of dense cores in \filAname\ is $\sim 20$\,\msun, 
 around 10 times higher than that in Taurus/L1517. Assuming free-fall gas accretion onto dense cores along the filament, we roughly
calculate the infall velocity via the relation $V_{\rm infall} = \sqrt{GM*{\rm cos}\theta/r}$, where G is the gravitational constant, $r$ is the radius within which gas 
flows onto dense cores, adopted to be $1/4\lambda$ (0.33\,pc), and $\theta$ is the inclination angle of the filament with respect to the line of sight. 
As a result, the average mass $\sim 20$\,\msun\ yields $V_{\rm infall} = \sim 0.5$\,\vel\ for $\theta=0\degr$, and $\sim 0.4$\,\vel\ for $\theta=45\degr$,
both of which are around 4 times greater than the amplitude of the velocity oscillation in \filAname.
Note that these infall velocities  should be overestimated to be the accretion velocity of gas onto dense cores since neither 
gas accretion is purely free-fall in reality nor the infall commences at an infinite radius as assumed in the above equation (e.g., \citealt{vaz19}).
Keeping this uncertainty in mind, and given 
core formation through GI being at work in G350.5 (see above, and Paper\,I), we suggest that 
core-forming gas motions induced by GI could contribute in part to the observed velocity oscillation.
  However, another mechanism (e.g., see Model\,2 in Fig.\,\ref{fig-models}, and Sect.\,\ref{sec:mhd}) is still required to explain 
the discrepancy between the observations and idealized GI-induced core-formation models, i.e., the lack of a constant phase shift 
observed between the velocity and density distributions.

\subsection{Large-scale MHD-transverse wave propagating along the filament?}
\label{sec:mhd}
As mentioned in Sect.\,\ref{sec:core:form:grav:contr}, there could be a mechanism responsible for the observed relation between the velocity and density distributions (see Fig.\,\ref{fig-pvNH2}). 
We conjecture that this mechanism might be the MHD-transverse wave propagating along the filament, which can also produce a periodic velocity oscillation. Actually, this mechanism was reported both in observations and in simulations \egcite{nak08,stu16}. For example,  combining both observed spatial and velocity undulations and the helical B-field measurements in the Orion integral-shaped filament (ISF), 
\citet{stu16} suggested that repeated propagation of transverse waves through the filament are progressively digesting the material that formerly connected Orion A and B into stars in discrete episodes. In three-dimensional MHD simulations of star formation in turbulent, magnetized clouds, including feedback from protostellar outflows, \citet{nak08} mentioned that stellar feedback like outflows  can induce large-amplitude Alfv\'en waves, which perturb the
field lines in the envelope that thread other parts of the condensed sheet. The large Alfv\'en speed in the diffuse envelope allows different
parts of the sheet to interact with each other quickly. Such interaction can spring up global, magnetically mediated oscillations for the condensed material \citep{nak08}. The appearance of a large-scale MHD-transverse wave along the filament is understandable as long as there is a poloidal component of B-field, which is expected with generally helical and other configurations of 3D magnetic fields \egcite{hei97,fie00,fie00b,stu16,sch18,rei18}

In the filament \filAname, the driving source of the large-scale MHD-transverse wave could result from outflows of young stellar objects \citep{nak08}. In addition,
the gas accretion flows onto the filament (see Paper\,I) could be an additional driving source. 
 In principle, a wave-like shape of spatial density distribution could be expected as observed in the Orion-A ISF \citep{stu16}. However, it can not be recognized from Fig.\,\ref{fig-13CO_emission}. This could be because of projection effects. That is, the large-scale MHD-transverse wave oscillates toward and away from us (see model\,2 in Fig.\,\ref{fig-models}) while propagating along the filament. As a result, the filament is projected to be a rather straight morphology on the plane-of-sky but the periodic velocity oscillation can be observed to blue and red-shifted with respect to the systemic velocity. 

To conclude, given the GI-induced core formation being at work in \filAname\ (see Sect.\,\ref{sec:core:form:grav:contr}), we suggest that
the observed periodic velocity oscillation may result from a combination of 
the core-forming gas motions induced by GI 
and a periodic physical oscillation driven by a MHD transverse wave. This combination can make $\lambda/4$ shift, as expected between the velocity and density distributions in the core-formation models (see Model\,1 of Fig.\,\ref{fig-models}), disappear due to the mixture of the two different motions.
Instead, the wave could cause the correspondence 
between the velocity extremes and density enhancements  (see Fig.\,\ref{fig-pvNH2}) if the wave is stronger than the core-forming gas motions in velocity amplitude. 
Despite of no definitive interpretation, it is worthwhile to test the possibility of the large-scale transverse wave being at work in \filAname. Therefore, we call for 
future high sensitivity/resolution Zeeman, and dust polarization measurements to infer the B-field strength and to constrain the field morphology (see above).

\section{Conclusions}
We have analysed the internal kinematics of the filament \filname\ with
our observations of \thco, and \etco~(2-1) by APEX. \thco\ emission
reveals two clouds with different velocities. The major cloud \filname\ corresponds
to $-5.7$ to $-2$\,\vel\ while the other one corresponds to  $-9.0$ to $5.7$\,\vel.
Our analysis shows 
that the filament \filname\ as a whole is supersonic and gravitationally bound.
In addition, we find a large-scale periodic velocity oscillation along the filament \filAname\ with a wavelength of $\sim 1.3$\,pc and an amplitude of $\sim 0.12$\,\vel. Comparing with the gravitational-instability induced core formation models, we suggest that
the observed periodic velocity oscillation may result from a combination of the kinematics of gravitational instability-induced core formation and a periodic physical oscillation driven by a MHD transverse wave. To test the latter, future high sensitivity, and resolution Zeeman, and dust polarization measurements toward \filname\ are required to infer the B-field strength and to constrain the field morphology

\medskip
\noindent{\textbf{Acknowledgments}}\\
We thank the anonymous referee for constructive comments that improved the quality of our paper. 
This work was in part sponsored by the Chinese Academy of Sciences (CAS), through a grant to
the CAS South America Center for Astronomy (CASSACA) in Santiago, Chile.
  AS acknowledges funding through Fondecyt regular (project code
1180350), ``Concurso Proyectos Internacionales de Investigaci\'on''
(project code PII20150171), and Chilean Centro de Excelencia en
Astrof\'isica y Tecnolog\'ias Afines (CATA) BASAL grant AFB-170002.  J. Yuan is supported by the National Natural Science Foundation of China
through grants 11503035, 11573036.
 This research made use of Astropy,
a community-developed core Python package for Astronomy (Astropy
Collaboration, 2018).

\bibliographystyle{mnras}
\bibliography{../paper}
%
\appendix
\section{Optical depth}
\label{sec:opt:dep}
\begin{figure}
\centering
\includegraphics[width=3.4 in]{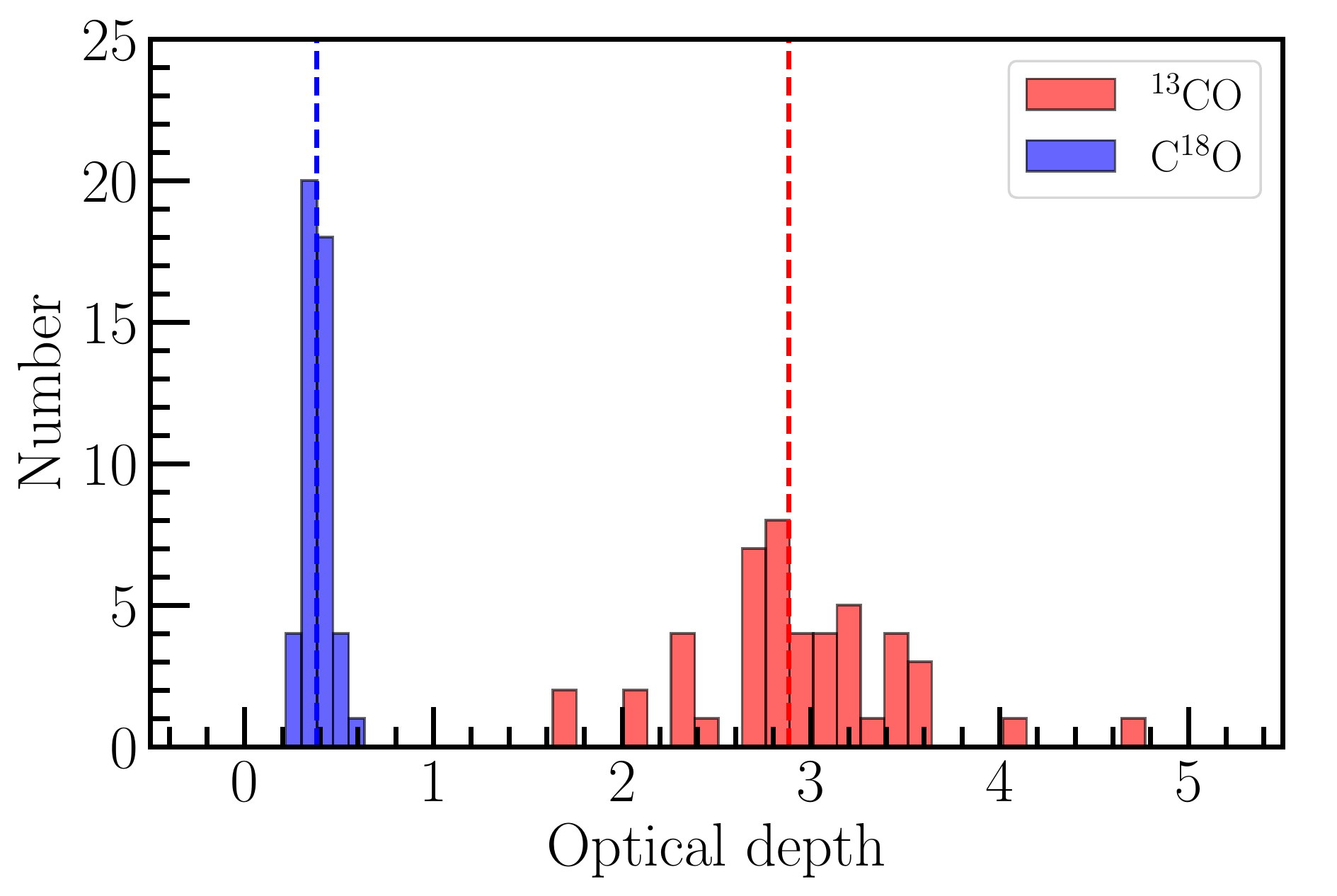}
\caption{Histogram of the optical depth for \thco\ (a) and \etco\ (b). The majority of \etco~(2-1) emission appears to be optically thin while \thco~(2-1) emission
suffers more optically thick effects.
}
\label{fig-opt_depth}
\end{figure}

Following the radiative transfer equations with an assumption of 
optically thin \etco~(2-1) emission \egcite{gar91}, we calculated the optical depth ($\tau$) of \thco, and \etco~(2-1) for 
the selected positions in the main structure of \filname\ as below:
\begin{equation}\label{eq:luminosity}\label{eq:plummer}
\frac{T_{\rm mb}^{{{\rm ^{13}CO}}}}{T_{\rm mb}^{{{\rm C^{18}O}}}} \approx \frac{1-e^{\tau_{13}}}{1-e^{\tau_{18}}} \approx \frac{1-e^{\tau_{13}}}{1-e^{\tau_{13}/R}},
\end{equation} 
where $T_{\rm mb}$ is the main-beam temperature for the two species, and the isotope ratio $R = [{\rm ^{13}CO}]/[{\rm C^{18}O}]$ is adopted to be 7.7 following the derivation in \citet{yua16}. 
Instead of the weak-emission component of \thco~(2-1), its main velocity component matching the \etco~(2-1) counterpart was taken into account in the practical 
calculations. With $\tau_{13}$ estimated, we can obtain $\tau_{18}$ via the relation $\tau_{18}=\tau_{13}/R$.
The statistics of the estimated optical depths for both species is shown in Fig.\,\ref{fig-opt_depth}. It can be seen that
 all of \etco~(2-1) emission in the main structure of \filname\ are optically thin while \thco~(2-1) emission is optically thick with optical depths up to $4.6$. 

\section{Excitation temperature}
\label{sec:ex:temp}
We further evaluated the excitation temperatures for
the selected positions in the main structure of the filament following \citet{liu15}:

\begin{equation}\label{eq2}
\begin{array}{rrr}
T_{\rm mb}=\frac{\mathrm{h} \nu}{\mathrm{k}}[\frac{1}{J_{\nu}(T_{\mathrm{ex}})}-\frac{1}{J_{\nu}(T_{\mathrm{bg}})}]
$$[1-\mathrm{exp}(-\tau)]f,
\end{array}
\end{equation}
where $J_{\nu}$ is defined as $\frac{1}{\mathrm{exp}(\mathrm{h} \nu/\mathrm{k}T)-1}$, $T_{\rm mb}$ is the main-beam temperature,
 $T_{\mathrm{ex}}$ is the exciting temperature, and $T_{\mathrm{bg}}=2.73$ K is the cosmic background radiation;
$\tau$ is the optical depth,  the fraction of the telescope
beam filled by emission $f$ is assumed to be 1, B and $\mu$ are the rotational constant and the permanent dipole moment of molecules
respectively. Given the optical depth (in Sect.\,\ref{sec:opt:dep}) and $T_{\rm mb}^{{{\rm ^{13}CO}}}$, we obtained the excitation temperature of
\thco~(2-1) from Eq.\,\ref{eq2}. As a result, the average excitation 
temperature of \thco~(2-1) in the main structure of the filament is $12\pm0.8$\,K, which is $\sim 6$\,K less than the corresponding 
average dust temperature there derived from the dust temperature map (Paper\,I). 
This difference may be related to non full thermalization of \thco~(2-1) in the main structure of \filname.

\section{Gaussian fitting results}
\begin{figure*}
\centering
\includegraphics[width=6.8 in]{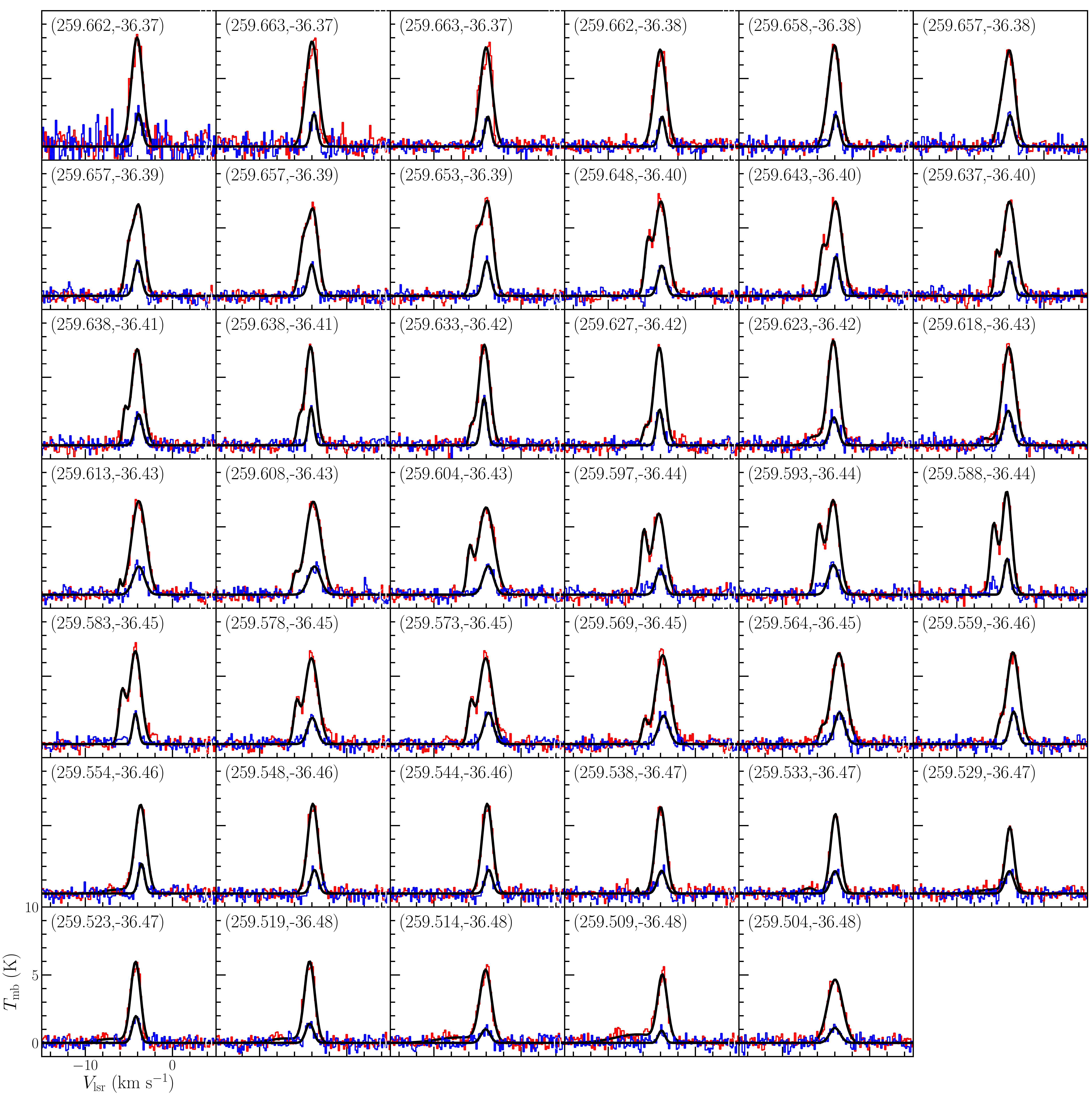}
\includegraphics[width=6.8 in]{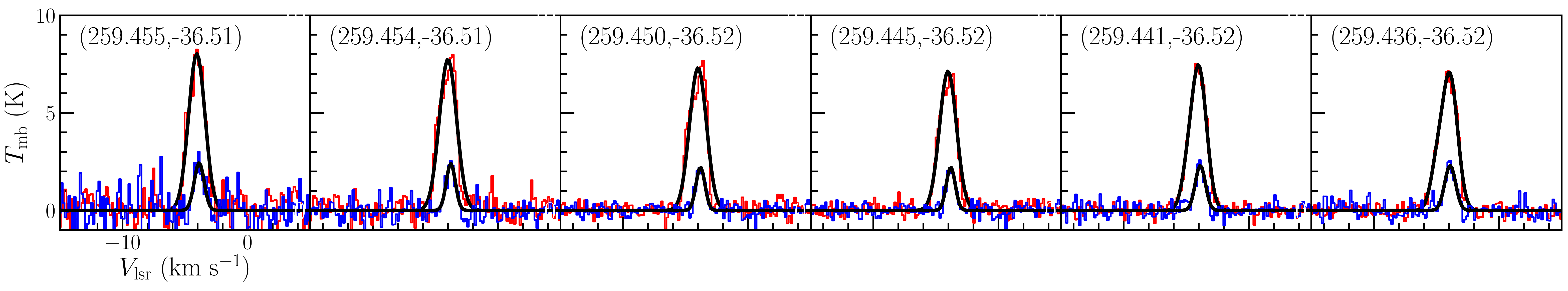}
\caption{Gaussian-fitting results of both \thco\ and \etco~(2-1) for the selected 41 positions in \filAname\ (upper panel), and 6 positions in \filBname\ (bottom panel).
The red and blue colors represent the observed spectra of \thco\, and \etco~(2-1), respectively. The black stands for the fitting result. 
}
\label{fig-guassA}
\end{figure*}
\bsp
\label{lastpage}
\end{document}